\documentclass[11pt]{article}
\usepackage{moriond,epsfig}

\def\bq{\begin{eqnarray}}
\def\eq{\end{eqnarray}}
\def\l{\langle}
\def\r{\rangle}

\begin{document}
\begin{flushright}
  UPRF-2002-06
\end{flushright}
\vspace*{4cm}
\title{NLO PARTONIC EVENT GENERATORS}

\author{ S. WEINZIERL }

\address{Dipartimento di Fisica, Universit\`a di Parma,\\
INFN Gruppo Collegato di Parma, 43100 Parma, Italy}

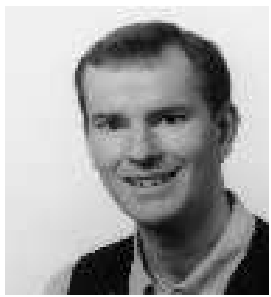
\maketitle\abstracts{
Merging the benefits of an NLO calculation with event generators
is a topic of current interest.
Among other things, it is advantageous to be able to generate unweighted
events according to NLO matrix elements.
I report on an algorithm which generates a sequence of unweighted
momentum configurations, each configuration containing either n or n+1
four-vectors, such that for any infrared safe observable the average
over these configurations coincides with the NLO calculation up to errors
of a resolution variable.}

\section{Introduction}

As high energy experiments move towards a new generation of colliders,
accurate simulation of collision events becomes increasingly important.
In particular this is the case for QCD, which often constitutes the bulk 
of the events.
Theoretical predictions are in most cases either based on exact perturbative calculations,
resummations or event generators.
These methods have complementary regions of applicability.
Exact perturbative calculations give a prediction for an observable
at the parton level
as a power series in the coupling constant.
The state-of-the-art are numerical next-to-leading order programs, which
allow the calculation of any infra-red safe observable for a given process.
They give a reliable prediction, if all scales entering an observable
are of the same order of magnitude, but are limited to a few partons only.
If an observable depends on a large ratio of two scales, large logarithms
appear in the perturbative expansion.
In this situation one often reorganises the perturbative expansion
and resums the large logarithms to all orders.
Within resummation, the state of the art is next-to-leading logarithmic
accuracy.
Up to now, these resummations have to be carried out for each observable
individually.
Event generators follow a different philosophy and rely on a three-stage process:
Starting from a hard-scattering process, parton showers are generated in the
second stage. Finally, a hadronization model converts partons into hadrons.
The first two stages, hard scattering and parton showering, can be described in perturbation
theory, while hadronization involves long-distance physics
and is non-perturbative.
Event generators usually only employ leading-order matrix elements and partons
showers accurate to leading logarithmic accuracy.
They describe reasonably well multiple collinear emission
and they generate unweighted events,
since they only use leading-order matrix elements.
However, they do a rather poor job for multi-jet events.
This is due to the fact that they only use the simplest $2 \rightarrow 2$
and $2 \rightarrow 3$ LO matrix elements, that the showering algorithm
does not populate the complete phase space and that color interference
is only approximated by angular ordering.
Since multi-jet events become increasingly important for new physics searches,
it is a topic of current interest to merge the benefits
of exact perturbative calculations and event generators.
Several issues have to be addressed for an event generator according to NLO
matrix elements:
\begin{itemize}
\item Double counting: Gluon emission can originate from the hard scattering
as well as from parton showering. 
\item Color flow: Matrix elements are usually color summed, where as the algorithms
for showering and hadronization are based on open color lines.
\item Negative weights: A standard NLO program generates $n$- and $(n+1)$-parton
phase space contributions, which are finite, but not necessarily positive definite.
\end{itemize}
Recent developments in this field considered combining 
parton showers with multi-parton tree-level matrix elements 
$\!{}^{1-3}$,
with resummations of soft gluons
\cite{Mrenna:1999mq},
and with NLO matrix elements
$\!{}^{5-9}$.
In this talk I focus on the issue of negative weights.
Negative weights are not a problem per se, but can seriously waste computer
resources, if events are passed through a detector simulation.
It is generally the case that the CPU time needed for detector simulation exceeds
the one needed for the generation of an event.
It is therefore desirable to avoid them from the very beginning.
In standard NLO calculations
negative weights occur since individual contributions to
an observable are not necessarily positive definite.
It is therefore natural, to consider first a ``partonic event generator'',
leaving parton showers and hadronization for the moment aside.
Here I report on an algorithm which generates a sequence of unweighted
momentum configurations, each configuration containing either n or n+1
four-vectors, such that for any infrared safe observable the average
over these configurations coincides with the NLO calculation up to errors
of a resolution variable \cite{Weinzierl:2001ny}.
I focus on electron-positron annihilation with massless partons.
The algorithm is independent of the specific hard process and
can be implemented on top of existing NLO programs.
After an outline of the basic idea, I focus on a few technical challenges
which have to be addressed in order for the algorithm to 
run efficiently.

\section{The algorithm}

Before stating the algorithm I briefly recall the main features of
standard NLO calculations.
The calculation of an observable at NLO receives contributions from
the Born matrix element, the virtual corrections and the real emissions.
Taken separately, the last two contributions are divergent, only their
sum is finite.
The subtraction- or phase space slicing method can be used to render them
finite individually.
Within the subtraction method one adds and subtracts a suitable piece $d\sigma^A$ \cite{Catani:1997vz}:
\bq
\sigma^{NLO} = \int\limits_n d\sigma^B 
 + \int\limits_n \left( d\sigma^V + \int\limits_1 d\sigma^A \right) 
 + \int\limits_{n+1} \left( d\sigma^R - d\sigma^A \right)
\eq
The modified virtual corrections and the modified real emissions are
now separately finite, but not necessarily positive definite.
The basic idea to generate unweighted events according to NLO
matrix elements is relatively simple 
and inspired by a slicing approach (although it will entirely based on the subtraction method): 
Consider a $(n+1)$-parton configuration. If all
partons are separated by at least $y_{res}$ according to a resolution criteria, one returns this $(n+1)$-parton
configuration.
If on the other hand a parton is not resolved at a scale $y_{res}$, one averages
over the real emissions up to the scale $y_{res}$, combines the result with 
the virtual corrections
and returns a $n$-parton configuration.
By choosing ${y_{res}}$ large enough the sum will have a positive weight.
For an efficient implementation I would like to discuss three technical points:
\begin{itemize}
\item A modified version of the subtraction method, such that the 
subtraction terms are non-vanishing only for ${y < y_{res}}$.
\item A parameterization of the phase space, which starts
from a ${n}$ parton configuration and inserts an additional soft or collinear
particle, such that the integration over the unresolved part involves only a three-dimensional integration
and, in addition, such that the dominant ${1/s_{ij}}$-behaviour of the integrand is absorbed into the measure.
This ensures that the unresolved region can be integrated efficiently.
\item The Metropolis algorithm to generate events with constant weight 
according to a (complicated) probability distribution.
\end{itemize}
I start with a modified version of the subtraction method:
The subtraction method has in general the advantage that it does not
introduce additional approximations to the matrix elements or to the phase space.
The only approximation made for the algorithm presented here,
consists in replacing a $(n+1)$-parton configuration
by a $n$-parton configuration, whenever two partons are not resolved within
$y_{res}$. 
In order for the basic idea outlined above to work, nothing should be subtracted if all
partons are separated by $y_{res}$.
In the original version of the dipole formalism \cite{Catani:1997vz} 
the approximation term is given by a sum over dipoles
$\sum {\cal D}_{ij,k}$
and subtracted over the complete phase space.
However, it turns out that the approximation term can still 
be integrated analytically if the region where it is
subtracted, is restricted \cite{Nagy:1998bb}:
\bq
d \sigma^A & = &\sum {\cal D}_{ij,k} \Theta \left(y_{ij,k} < y_{res} \right) 
\eq
Here,
$y_{ij,k} = 2p_ip_j/(2p_ip_j+2p_ip_k+2p_jp_k)$
is the appropriate resolution variable if partons $i$ and $j$ become close
to each other and $k$ is the spectator.\\
The second point is a suitable parametrization of the phase space.
For any given ``hard'' $n$-parton configuration one needs a parametrization
for the insertion of an additional parton, which can become soft or collinear
to one of the $n$ hard partons.
This is done as follows \cite{Weinzierl:1999yf}:
The phase space measure of the $(n+1)$-parton configuration can be related
to a $n$-parton configuration by
\bq
d\phi_{n+1} = d\phi_n \frac{ds_{as} ds_{sb} d\phi_s}{4 (2\pi)^3 s_{ab}}.
\eq
The invariants $s_{as}$, $s_{sb}$ and the angle $\phi_s$ are parametrized
in terms of three random variables $v_1$, $v_2$ and $v_3$, uniformly
distributed in $[0,1]$:
\bq
s_{as} = s_{ab} \left( \frac{s_{min}}{s_{ab}} \right)^{v_1},
s_{sb} = s_{ab} \left( \frac{s_{min}}{s_{ab}} \right)^{v_2},
\phi_s = 2 \pi v_3.
\eq
$s_{min}$ is an internal technical parameter, which is chosen to be small.
It is possible to reconstruct the four-vectors of the ${n+1}$ parton 
configuration from these invariants.
The phase space measure becomes
\bq
d\phi_{n+1} = d\phi_n \frac{1}{4 (2\pi)^2} \frac{s_{as}s_{sb}}{s_{ab}} 
 \ln^2\left( \frac{s_{min}}{s_{ab}} \right) dv_1 dv_2 dv_3
\eq
and the 
${1/s_{ij}}$-behaviour of the integrand has been absorbed into the measure.
\\
With an appropriate choice of $y_{res}$ one obtains now events with positive
weights.
In general, the probability distribution will be a rather complicated function
and unweighting according to the standard acceptance-rejection method
might not be feasible.
The method of choice is therefore the Metropolis algorithm, which allows
to generate unweighted events 
according to a multi-dimensional probability density
${P(x_1,...,x_d)}$, which not necesarrily factorizes.
One starts from a state ${\vec{x} = (x_1,...,x_d)}$
and randomly generates a new candidate ${\vec{x}'}$.
One then calculates the quantity${\Delta S = - \ln(P(\vec{x}')/P(\vec{x}))}$
and accepts the new candidate with probability ${\mbox{min}(1,e^{-\Delta S})}$.
The way new candidates are suggested is arbitrary, restricted only by 
ergodicity and detailed balance:
Each state must be reachable within a finite number of steps and 
the probability of suggesting the state ${x_2}$ given we are in state ${x_1}$
has to be equal to the probability of suggesting ${x_1}$ given we are in ${x_2}$.
\\
To summarize, the complete algorithm to generate unweighted events according to NLO matrix elements
is therefore as follows:
One first picks a set of random numbers ${u_1,...,u_m,x,v_1,v_2,v_3}$,
where ${u_1,...,u_m}$ define the ``hard'' n-parton configuration,
${x}$ decides where the ``soft'' particle gets inserted and
${v_1,v_2,v_3}$ (together with the  ``hard'' configuration and the information where the ``soft''
particle gets inserted) yield the ``soft'' ${(n+1)}$ event.
One then checks for the $(n+1)$-parton event if all $y_{ij,k}$ are greater than $y_{res}$, e.g. if all 
partons are well separated.
If this is the case the probability $P'$ is proportional to the real emission matrix element
and the $(n+1)$ parton configuration will be returned, if the event is accepted by the Metropolis
algorithm.
If not all partons are well separated, the ``hard'' $n$-parton configuration has the probability
\bq
P' & \sim & \left| {\cal M}_{Born}^{n} \right|^2 
       +  2 \; \mbox{Re} \; {{\cal M}_{Born}^{n}}^\ast {\cal M}_{1-loop}^{n} 
       + \l {\cal M}_{Born}^{n} \left| {\bf I} \right| {\cal M}_{Born}^{n} \r \nonumber \\
 & & + \int dv_1 dv_2 dv_3 \; w_{soft} \left( \left| {\cal M}_{Born}^{n+1} \right|^2 - \sum {\cal D}_{ij,k} \Theta \left(y_{ij,k} < y_{res} \right) \right)
\eq
The last terms in the first and second line are the added and subtracted dipole terms, which ensure
that the expressions over the $n$- and $(n+1)$-parton phase space are finite separately.
The second line involves a three-dimensional integration over the unresolved phase space.
In both cases, e.g. if the event is classified as a $n$- or $(n+1)$-parton event,
the new candidate gets accepted with Metropolis probability
${\mbox{min}(1,P'/P)}$, where $P'$ is the probability of the new candidate and $P$
is the probability of the previous event.

\section{Outlook}

I presented an algorithm to generate unweighted partonic events according to NLO
matrix elements in electron-positron annihilation.
An important ingredient is an efficient integration over the unresolved region
on a point-by-point basis in hard phase space.
The extension to initial-state partons or massive partons should be feasible.

\section*{References}


\end{document}